\documentclass[conference]{IEEEtran}

\usepackage{graphics}
\usepackage{epstopdf}
\usepackage[pdftex]{graphicx}
\usepackage{stfloats}
\usepackage{array}
\usepackage{float}
\usepackage[cmex10]{amsmath}
\usepackage{amsmath,amsfonts,amssymb}
\usepackage{graphicx,graphics}
\usepackage{enumerate}
\usepackage{color}
\usepackage{verbatim}
\usepackage{amsthm}
\usepackage{multirow}
\usepackage{subcaption}
\usepackage{siunitx}
\usepackage{amsmath,amsthm}
\usepackage{cleveref}
\usepackage[font=footnotesize]{caption}
\usepackage{fixltx2e}
\usepackage[USenglish]{babel}

\usepackage{geometry}

\usepackage{fancyhdr}

\newcommand{\e}{\mathrm{e}}

\begin{document}

\title{Practical Implementation of Adaptive Analog Nonlinear Filtering For Impulsive Noise Mitigation}

\author{
Reza Barazideh$^{\dag}$, Alexei V. Nikitin$^{\dag,\ast}$, Balasubramaniam Natarajan$^{\dag}$ \\
\small $^{\dag}$ Department of Electrical and Computer Engineering, Kansas State University, Manhattan, KS, USA.\\
$^{\ast}$ Nonlinear Corp., Wamego, KS 66547, USA.\\
Email:\{rezabarazideh,bala\}@ksu.edu, avn@nonlinearcorp.com}

\maketitle
\thispagestyle{fancy}

\begin{abstract}
It is well known that the performance of OFDM-based Powerline Communication (PLC) systems is impacted by impulsive noise. In this work, we propose a practical blind adaptive analog nonlinear filter to efficiently detect and mitigate impulsive noise. Specially, we design an Adaptive Canonical Differential Limiter (ACDL) which is constructed from a Clipped Mean Tracking Filter (CMTF) and Quartile Tracking Filters (QTFs). The QTFs help to determine a real-time range that excludes outliers. This range is fed into the CMTF which is responsible for mitigating impulsive noise. The CMTF is a nonlinear analog filter and its nonlinearity is controlled by the aforementioned range. Proper selection of this range ensures the improvement of the desired signal quality in impulsive environment. It is important to note that the proposed ACDL behaves like a linear filter in case of no impulsive noise. In this context, the traditional matched filter construction is modified to ensure distortionless processing of the desired signal. The performance improvement of the proposed ACDL is due to the fact that unlike other nonlinear methods, the ACDL is implemented in the analog domain where the outliers are still broadband and distinguishable. Simulation results in PRIME (OFDM-based narrowband PLC system) demonstrate the superior BER performance of ACDL relative to other nonlinear approaches such as blanking and clipping in impulsive noise environments. 

\end{abstract}


\section{Introduction}

 With the pervasive reach of powerline infrastructure, low deployment costs, and its wide frequency band, powerline communication (PLC) has become a strong candidate for a variety of smart grid applications \cite{Galli2011ForTheGrid}. High speed communication over powerlines has recently attracted considerable interest and offer a very interesting alternative to wireless communication systems. The ability to support high data rates in PLC requires multicarrier protocols such as orthogonal frequency division multiplexing (OFDM) \cite{Zhidkovn08_Simpleanalysis}. The two major issues in OFDM-based PLC are: (i) impedance mismatch that is due to the fact that the powerline infrastructure is originally designed for power delivery and not for communications \cite{Galli2011ForTheGrid}, and (ii) noise that typically consists of two parts: the thermal noise, which is assumed to be additive Gaussian noise, and impulsive noise that may be synchronous or asynchronous relative to the main frequency \cite{Standard,AdaptiveNoiseMitigation-2010}. Since OFDM employs a larger symbol duration (i.e., narrowband subcarriers), the energy of impulsive noise is naturally spread over all subcarriers. While this provides some level of robustness to impulsive noise, system performance can still degrade if impulse noise power exceeds a certain threshold \cite{ghosh1996}.

 A plethora of techniques to mitigate the effect of impulsive noise have been proposed over the past few decades. For example, channel coding techniques such as turbo codes (TC) \cite{Turbo2014Umehara} and low density parity check codes (LDPC) \cite{LDPC2005Nakagawa} have been used to improve bit error rate (BER) in  impulsive noise environments. It has been shown that these approaches are effective only in single carrier schemes and there is small gain in OFDM systems which are widely used almost in all PLC applications \cite{AdaptiveNoiseMitigation-2010}. The reduction of signal-to-noise ratio (SNR) in highly impulsive noise environments such as PLC can be too severe to be handled by forward error correction (FEC), frequency-domain block interleaving (FDI) \cite{Nassar12local} or time-domain block interleaving (TDI) \cite{Time_Interleaving}. Many approaches assume a statistical model such as $\alpha$-stable \cite{ Samorodnitsky94stable} and Middleton class A, B and C \cite{ Middleton99nongaussian} for the impulsive noise and use parametric methods in the receiver to mitigate impulsive noise. Such parametric methods require the overhead of training and parameter estimation. In addition, difficulty in parameter estimation and model mismatch degrade the system performance in non stationary noise.
The non-Gaussian nature of impulsive noise has also motivated the use of various memoryless nonlinear approaches such as clipping \cite{Tseng-2012-robust-clipping}, blanking \cite{Blanking}, joint blanking-clipping \cite{Blanking-Clipping}, linear combination of blanking and clipping \cite{LinearComBLNCLP2016Juwono}, and deep clipping \cite{DeepCliping_2014_Juwono}. As shown in \cite{Zhidkovn08_Simpleanalysis}, these methods have good performance only for asynchronous impulsive noise in high signal-to-impulsive noise ratios (SIR) and their performance degrades dramatically in severe impulsive environment. To address the challenge of severe impulsive noise conditions, a two-stage nulling algorithm based on iterative channel estimation is proposed in \cite{Two_Stage_Iterative} which is computationally intensive.

The current state-of-art approach to mitigate the effects of impulsive noise is to convert the analog signal to digital and then using digital nonlinear methods. This classical approach has two main problems. First, the signal bandwidth decreases in the process of analog-to-digital conversion and an initially impulsive broadband noise will appear less impulsive making it challenging to remove outliers via digital filters \cite{Nikitin-2011b-EURASIP}-\cite{Nikitin2015}. Although, this problem can be overcome by increasing the sampling rate (and thus the acquisition bandwidth), it exacerbates the memory, complexity and computational load on digital signal processor (DSP). Second, digital nonlinear filters are not ideally suited to real-time processing relative to analog filters due to added computational burden. Therefore, in our prior work we proposed a blind adaptive analog filter, referred to as Adaptive Nonlinear Differential Limiter (ANDL) to mitigate impulsive noise in analog domain before the analog-to-digital converter (ADC) \cite{Alexi-Dale}, \cite{Khodam_Latincom}. In \cite{Alexi-Dale}, we studied the basics of the ANDL approach and the general behavior of SNR in a conceptual system without realistic OFDM transmitter and receiver modules. In \cite{Khodam_Latincom}, we extended the analysis by explicitly qualifying the BER performance of the ANDL in a practical OFDM-based PLC system. Although, in \cite{Khodam_Latincom} a simple method is proposed to determine an effective value for the resolution parameter that maximizes the signal quality while mitigating the impulsive noise, finding the resolution parameter in real-time and practical implementation of the filter are still a open problem that we address in this paper.

In this paper, for the first time, a practical blind adaptive analog nonlinear filter, referred to as Adaptive Canonical Differential Limiter (ACDL) is proposed to mitigate the effect of impulsive noise in PLC system. In practice, the ACDL consists of two modules: (i) the range module that uses Quartile Tracking Filters (QTFs) to establish the range that excludes impulsive noise, and (ii) Clipped Mean Tracking Filter (CMTF) that consists of a nonlinear filter that mitigates outliers without knowledge of the noise distribution. The effects of this filter on the desired signal are totally different relative to that on the impulsive noise because of nonlinearity of this filter. Therefore, SNR in the desired bandwidth will increase by reducing the spectral density of non-Gaussian noise without significantly affecting the desired signal. We validate the performance of the ACDL by measuring the SNR and the BER of a practical PLC system. In addition, we highlight the preference of our approach rather than other conventional approaches such as blanking, clipping and linear filtering. 

The remainder of this paper is organized as follows. Section \ref{sec:System Model} describes the system and noise models. Section III details the proposed analog nonlinear filter modules and their practical implementation. Section IV presents simulation results and finally conclusions are drawn in Section V.

\section{System Model}\label{sec:System Model}

\begin{figure*}
\centering
\includegraphics[scale=.38]{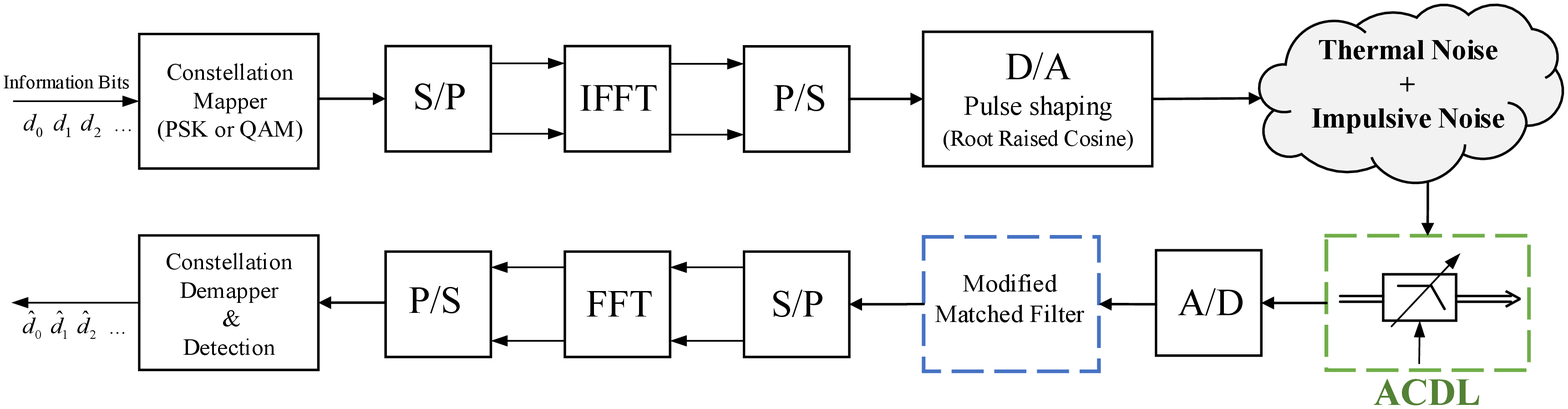}
\caption{System model block diagram.}
\label{fig:System Model}
\end{figure*}

The OFDM-based PLC system considered in this work is shown in Fig.~\ref{fig:System Model}. In this system, information bits are first modulated by phase shift keying (PSK) or quadrature amplitude modulation (QAM) schemes. The modulated data $s_k$ are passed through an inverse discrete Fourier transform (IDFT) to generate OFDM symbols over orthogonal subcarriers and then shaped by a root raised cosine waveform with roll-off factor $0.25$ and transmitted through the channel. The transmitted analog signal envelope in time domain can be expressed as

\begin{equation}
s(t) = \frac{1}{{\sqrt N }}\sum\limits_{k = 0}^{N - 1} {{s_k}}\,{\e^{j\frac{{2\pi kt}}{{{T}}}}}p(t),\,\,\,\,\,0 < t < {T},
\end{equation}
where $N$ is the number of subcarriers; $T$ is the OFDM symbol duration; and $p(t)$ denotes the pulse shape.
In general, for different applications, we can construct an OFDM symbol with $M$ non-data subcarriers and $N-M$ data subcariers. The non-data subcarriers are either pilots for channel estimation and synchronization, or null for spectral shaping and inter-carrier interference reduction. Under perfect synchronization, the received signal in an additive noise channel is given by
\begin{equation} \label{eq:recived signal}
r(t) = s(t) + w(t) + i(t).
\end{equation}
Here, $s(t)$ denotes the desired signal with variance $\sigma _s^2$; $w(t)$ is complex Gaussian noise with mean zero and variance $\sigma _w^2$; and $i(t)$ represents the impulsive noise which is not Gaussian and it is assumed that $s(t)$, $w(t)$, and $i(t)$ are mutually independent. In general, the model in \eqref{eq:recived signal} can be expanded to include channel attenuation (fading) effect. However, since the goal of the work is to demonstrate a novel approach to mitigation of impulsive noise, we restrict ourselves to additive noise channel model in \eqref{eq:recived signal}. It is important to note that the proposed ACDL approach is applicable to alternate channel model as well. As shown in Fig.~\ref{fig:System Model}, the conventional structure of the receiver is modified in order to deal with impulsive noise $i(t)$ and the additional proposed filter is implemented before the ADC as a front end filter.
Non-Gaussian noise $i(t)$ is the main challenge in the PLC and it has been shown by field measurements that cyclostationary impulsive noise and asynchronous impulsive noise are dominant in narrowband PLC (NB-PLC) and broadband PLC (BB-PLC), respectively \cite{Lin13impulsive_SparseBayesian}. Since both types of impulsive noises are presented in the NB-PLC \cite{Lin13impulsive_SparseBayesian, Alexi-Dale}, we consider both of them simultaneously. Based on field measurements \cite{Standard}, the dominant part of cyclostationary impulsive noise is a strong and narrow exponentially decaying noise burst that occurs periodically with half the alternate current (AC) cycle. As discussed in our previous work \cite{Khodam_Latincom}, this noise corresponds to
\begin{equation}\label{CS}
i_{\rm cs}(t) = A_{\rm cs}\, \nu (t) \sum\limits_{k = 1}^\infty  \exp \left(\! \frac{ - t \!+\! \frac{k}{2f_{\rm AC}}}{\tau _{\rm cs}} \!\!\right) \theta \left(t \!-\! \frac{k}{2f_{\rm AC}} \right),
\end{equation}
where $A_{cs}$ is a constant; $\tau_{cs}$ is characteristic decay time for the cyclostationary noise; $\nu(t)$ is complex white Gaussian noise process with zero mean and variance one; and $\theta (t)$ is the Heaviside unit step function. The spectral density of this noise is shaped based on measured spectrum of impulsivity in practice (power spectral density (PSD) decaying at an approximate rate of 30 dB per 1 MHz) \cite{Standard}. On the other hand, asynchronous impulsive consists of short duration and high power impulses with random arrival. According to \cite{Khodam_Latincom}, this noise can be modeled as
\begin{equation}\label{AS}
i_{\rm as}(t) = \nu (t) {\sum\limits_{k = 1}^\infty  {A_k}\, \theta (t - {t_k})\, \e^{\frac{{ - t + {t_k}}}{\tau _{\rm as}}}}\,,
\end{equation}
where $A_k$ is the amplitude of ${k^{th}}$ pulse; $t_k$ is a arrival time of a Poisson process with parameter $\lambda$; and  $\tau_{as}$ is characteristic decay time for the asynchronous noise and has a duration about few microseconds. In the next section, we discuss the ACDL design and implementation in detail.

\begin{figure*}
\centering
\includegraphics[scale=.60,angle=0]{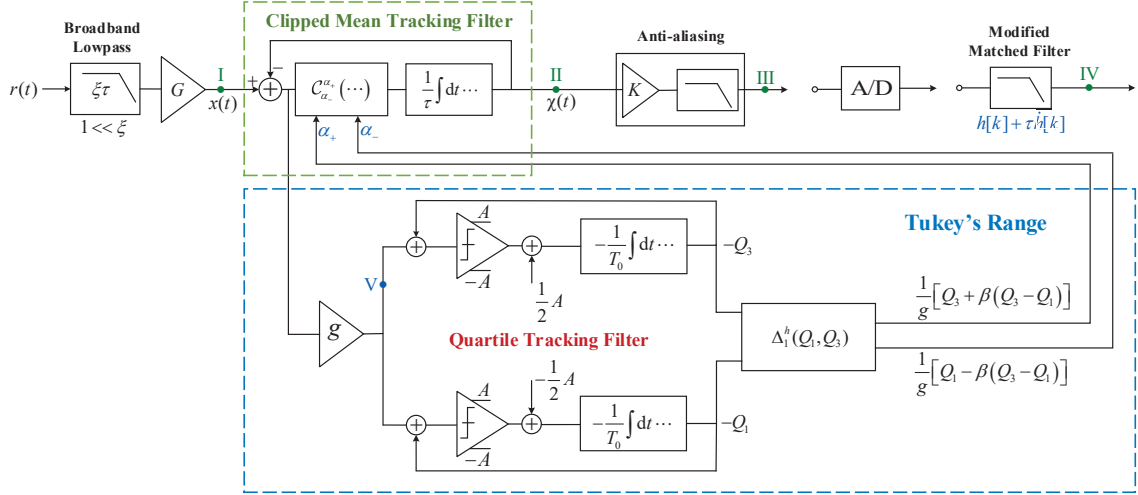}
\caption{Practical implementation of ACDL.}
\label{fig:Practical_Imp}
\end{figure*}

\section{Practical Implementation of ACDL }

The principal block diagram of the ACDL is shown in Fig.~\ref{fig:Practical_Imp}. Without loss of generality, it is assumed that the output ranges of the active components (active filters, integrators, and comparators), as well as the input range of the ADC, are limited to a certain finite range, e.g., to the power supply range $\pm V_c$. The time parameter $\tau$ is such that $1/{2\pi\tau}$ is equal to the corner frequency of the anti-aliasing filter (e.g., approximately twice the bandwidth of the signal of the interest $B_x$), and the time constant $T_0$ is two-to-three orders of the magnitude larger than $B_x^{ -1}$. The purpose of the front-end lowpass filter is to limit the input noise power and at the same time its bandwidth should remain sufficiently wide (i.e., $\xi\gg 1$), so that the impulsive noise is not excessively spread out in time. In general, we can assume that the gain $K$ is constant and is largely depended on the value of the parameter $\xi$ (e.g., $K\sim \sqrt \xi$ ), and the gains $G$ and $g$ are adjusted in order to fully utilise the available output ranges of the active components, and the input range of the ADC. For instance, $G$ and $g$ may be chosen to ensure that the average absolute value of the output signal (i.e., observed at point IV) is approximately $V_c/10$, and the difference $Q_3-Q_1$ is $2V_c/5$.

\begin{figure*}
\centering
\includegraphics[scale=.70,angle=0]{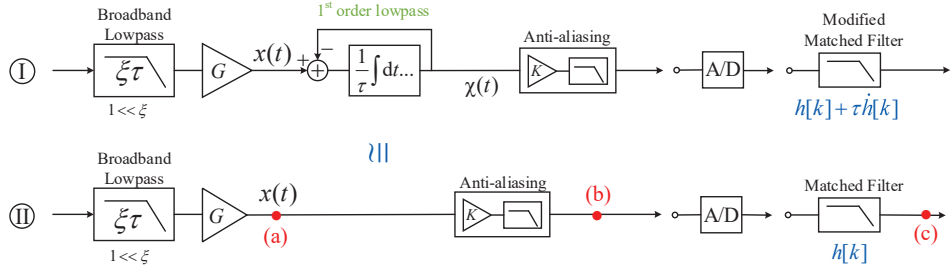}
\caption{Equivalent block diagram of Fig. \ref{fig:Practical_Imp} operating in linear regime.}
\label{fig:Linear}
\end{figure*}

\subsection{Clipped Mean Tracking Filter (CMTF)}

The role of the CMTF is to mitigate outliers from the input signal and at the same time it should be designed to behave like a linear filter in the absence of outliers. As shown in the block diagram of the CMTF in Fig.~\ref{fig:Practical_Imp}, the input $x(t)$ and the output $\chi(t)$ signals can be related by the following first order nonlinear differential equation
\begin{equation}\label{CMTF}
\frac{{\rm{d}}}{{{\rm{d}}t}}\chi(t)  = \frac{1}{\tau }\mathcal{C}_{\alpha_- }^{\alpha_+ }\left( {x(t) - \chi(t) } \right),
\end{equation}
where the clipping function $\mathcal{C}_{\alpha_- }^{\alpha_+ }(x)$ is defined as
\begin{equation}\label{Clipping}
\mathcal{C}_{\alpha_- }^{\alpha_+ }(x) = \left\{ \begin{array}{l}
 \alpha_+ \,\,\,\,for\,\,\,\,x >  \alpha_+  \\
 \alpha_- \,\,\,\,for\,\,\,\,x <  \alpha_- \\
\,\,x\,\,\,\,\,\,\,\,otherwise
\end{array} \right.,
\end{equation}
where $\alpha +$ and $\alpha -$ are the upper and lower clipping values, respectively. Note that for the clipping values such that $\alpha_- \leq x(t)-\chi(t)\leq \alpha_+ $ for all t, equation \eqref{CMTF} describes a first order linear lowpass filter with corner frequency $1/{2\pi\tau}$, and the filter shown in Fig.~\ref{fig:Practical_Imp} operates in a linear regime as shown in Fig.~\ref{fig:Linear}. However, when the values of the difference signal $x(t)-\chi(t)$ are outside of the interval $[\alpha_-,\alpha_+]$, the rate of change of $\chi(t)$ is limited to either $\alpha_-/\tau$ or $\alpha_+/\tau$ and no longer depends on the magnitude of $x(t)-\chi(t)$. Thus, if the values of the difference signal that lie outside of the interval $[\alpha_-,\alpha_+]$ are outliers, the output $\chi(t)$ will be insensitive to further increase in the amplitude of such outliers. In this work, an effective value of the interval $[\alpha_-,\alpha_+]$ is obtained as the \emph{Tukey's range} \cite{Tueky}, a linear combination of first ($Q_1$) and the third ($Q_3$) quartiles of the linear-regime difference signal
\begin{equation}\label{Tukey}
[\alpha_-,\alpha_+]=[Q_1-\beta(Q_3-Q_1),Q_3+\beta(Q_3-Q_1)],
\end{equation}
where $\beta$ is a constant coefficient (e.g., $\beta=3$). As illustrated in panel I of Fig.~\ref{fig:Linear}, in the linear regime the CMTF operates as a first order linear lowpass filter with time constant $\tau$. Then the quartiles $Q_1(t)$ and $Q_3(t)$ are obtained as output of the QTFs described in the next subsection.


\subsection{Quartile Tracking Filters (QTFs)}\label{sec:QTF}

\begin{figure}[t]
\centering
\includegraphics[width=.5\textwidth,height=50mm]{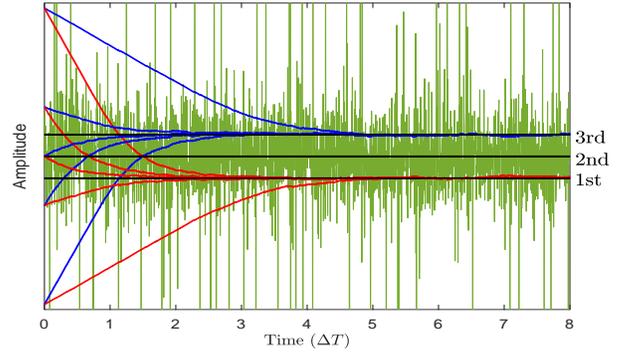}
\caption{Illustration of QTFs' convergence to steady state for different initial conditions. Eb/N0 = 0 dB, SIR = 0 dB.}
\label{fig:Quartile}
\vspace{-.25cm}
\end{figure}

Let $y(t)$ be a quasi-stationary bandpass (zero-mean) signal with a finite interquartile range (IQR), characterised by an average crossing rate $\langle f_0\rangle$ of the threshold equal to the third quartile of $y(t)$. (See \cite{Nikitin03signal} for discussion of quantiles of continuous signals, and \cite{Rice_CrossingRate} for discussion of threshold crossing rates.) Let us further consider the signal $Q_3(t)$ related to $y(t)$ by the following differential equation
\begin{equation}\label{eq:QTF_3}
\frac{{\rm{d}}}{{{\rm{d}}t}}{Q_3} = \frac{A}{T_0}\left[ {{\rm{sgn}}(y - {Q_3}) + \frac{1}{2}} \right],
\end{equation}
where $A$ is a constant (with the same units as $y$ and $Q_3$), and $T_0$ is a constant with the units of time. According to equation \eqref{eq:QTF_3}, $Q_3(t)$ is a piecewise-linear signal consisting of the alternating segments with positive $(3A/(2T_0))$ and negative $(-A/(2T_0))$ slopes. Note that $Q_3(t)\approx \rm{const}$ for a sufficiently small $A/T_0$ (e.g., much smaller than the product of the IQR and the average crossing rate $\langle f_0\rangle$ of $y(t)$ and its third quartile), and a steady-state solution of equation \eqref{eq:QTF_3} can be written implicitly as
\begin{equation}
\overline {\theta \left( {{Q_3} - y} \right)}  \approx \frac{3}{4},
\end{equation}
where the over-line denotes averaging over some time interval $\Delta t \gg \langle f_0\rangle^{-1}$. Thus, $Q_3$ approximate the \emph{third quartile} of $y(t)$ in the time interval $\Delta t$. Similarly, for
\begin{equation}\label{eq:QTF_1}
\frac{{\rm{d}}}{{{\rm{d}}t}}{Q_1} = \frac{A}{T_0}\left[ {{\rm{sgn}}(y - {Q_1}) - \frac{1}{2}} \right],
\end{equation}
a steady-state solution can be written as
\begin{equation}
\overline {\theta \left( {{Q_1} - y} \right)}  \approx \frac{1}{4},
\end{equation}
and thus $Q_1$ would approximate the \emph{first quartile} of $y(t)$ in the time interval $\Delta t$. Fig.~\ref{fig:Quartile} provides an illustration of the QTFs' convergence to the steady state for different initial conditions. In Fig.~\ref{fig:Quartile} signal $y(t)$ plotted by green line, first (red line) and third (blue line) quartiles, in comparison with the exact quartiles of $y(t)$ computed in the full shown time interval (black lines).

%

\subsection{Matched filter Modification}

\begin{figure}[t]
\centering
\includegraphics[width=.5\textwidth,height=60mm]{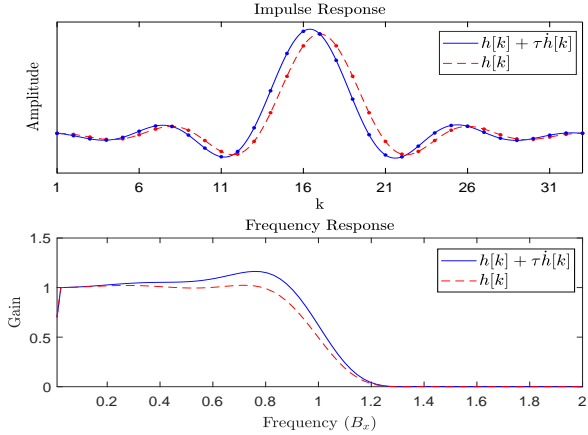}
\caption{Impulse and frequency response of matched filter and modified matched filter.}
\label{fig:MMF}
\end{figure}

\begin{figure}[t]
\centering
\includegraphics[width=.5\textwidth,height=60mm]{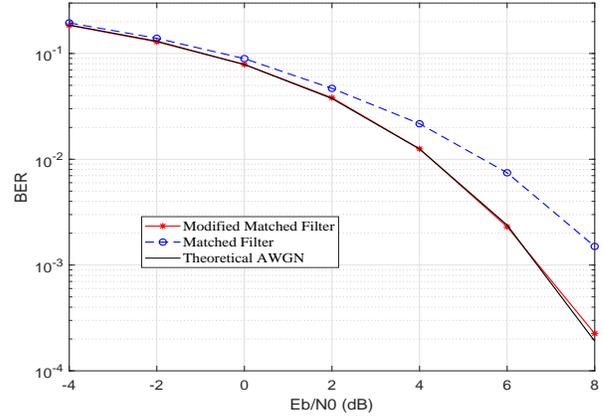}
\caption{Performance comparison between matched filter and modified matched filter in the presence of CMTF for BPSK modulation.}
\label{fig:MMFvsMF}
\end{figure}

In the absence of the CMTF in the signal chain, the matched filter (MF) following the ADC would have the impulse response $h[k]$ that can be viewed as a digitally sampled continuous-time impulse response $h(t)$ as shown in panel II of Fig.~\ref{fig:Linear}. Since our proposed filter should not have any negative impact when there is no impulsive noise, it is essential to modify the MF to compensate for the CMTF in a linear chain. We proposed a modification in the digital domain because it is simpler and does not need any extra components. In case of no impulsive noise CMTF acts like a first order linear lowpass filter with time constant $\tau$. Therefore, the relation between the input $x(t)$ and the output $\chi(t)$ can be expressed as
\begin{equation}\label{eq:CMTF_Lin}
x(t) = \chi(t)  + \tau \dot \chi(t).
\end{equation}
In the linear regime we want to have same result either with or without CMTF (Panels I and II in Fig.~\ref{fig:Linear}). Thus, the output of modified matched filter with the input $\chi(t)$ should be equal to the output of MF with the input $x(t)$. Therefore, we have
\begin{align} \label{eq:MMF} \notag
\chi(t) * h_{\rm mod}(t) &= x(t) * h(t)  \\
&=(\chi(t)  + \tau \dot \chi(t))*h(t),
\end{align}
where the asterisk denotes convolution and the impulse response $h_{\rm mod}[k]$ of the modified matched filter in the digital domain can be expressed as
\begin{equation}\label{eq:CMTF_Lin}
h_{\rm mod}[k] = h[k] + \tau \dot h[k].
\end{equation}
The impulse and frequency responses of the matched filter (a root-raised-cosine filter with roll-off factor $1/4$, bandwidth $5B_x/4$, and the sampling rate $8B_x$) and the modified matched filter (with $\tau=1/(4\pi B_x)$) are shown in Fig.~\ref{fig:MMF}. In the presence of CMTF the compensation of the modified matched filter on the BER performance of a OFDM system with $B_x=50$ kHz and BPSK modulation is shown in Fig.~\ref{fig:MMFvsMF}. As it can be seen the effect of CMTF in linear chain completely alleviated by the modified matched filter which means that our proposed filter does not harm the desired signal in case of no impulsive noise.

\section{Simulation results}\label{sec:Simulation results}

As a specific example, we simulate an OFDM-based PLC in accordance with the PRIME standard \cite{Prime-G3}. The sampling frequency is chosen as $f_s=250$ kHz and the FFT size is $N=512$, i.e., the subcarrier spacing $∆f=488$ Hz. As carriers 86–-182 are used for data transmission, the PRIME signal is located in the frequency range 42–-89 kHz \cite{Prime-G3}.
The system is studied in a noise environment and it consists of three components: (i) a thermal noise (ii) periodic cyclostationary exponentially decaying component with the repetition frequency at twice the AC line frequency ($2\times60$ Hz) and $\tau_{\rm cs}=200~\mu s$ (one tenth of OFDM symbol), and (iii) asynchronous random impulsive noise with normally distributed amplitudes captured by a Poisson arrival process with parameter $\lambda$ and $\tau_{\rm as}=2~\mu s$. Based on IEEE P1901.2 standard \cite{Standard} the PSD of noise components (i) and (ii) decay at a rate of 30 dB per 1 MHz. Since the cyclostationary noise is dominant in the NB-PLC, we set the power of this component three times higher than the asynchronous impulsive noise. To emulate the analog signals in the simulation, the digitization rate is chosen to be significantly higher (by about two orders of magnitude) than the ADC sampling rate. In the following, SNR and BER of an OFDM system with BPSK modulation are used as two metrics to evaluate the performance of the proposed analog nonlinear filter in comparison with other conventional approaches such as linear filtering, blanking and clipping.

\begin{figure}[t]
\centering
\includegraphics[width=.52\textwidth,height=96mm]{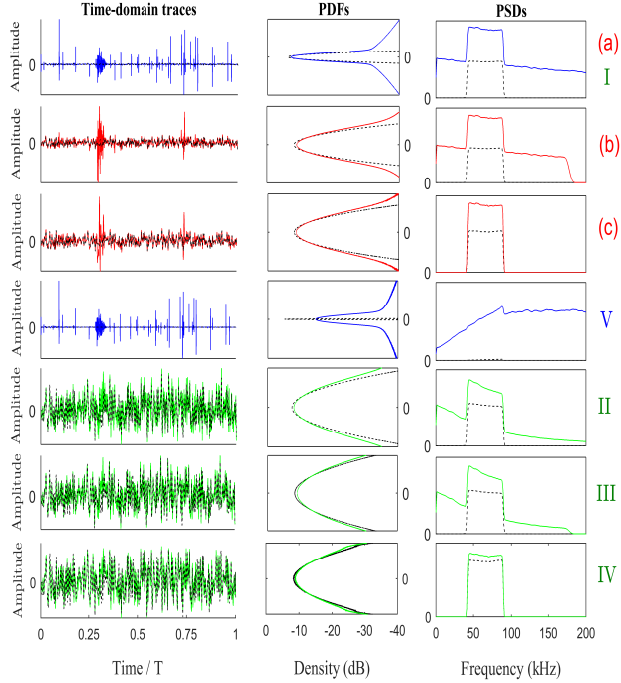}
\caption{Illustration of changes in the signal time- and frequency domain properties, and in its amplitude distribution. Eb/N0 = 10 dB, SIR = 1 dB.}
\label{fig:PSD}
\end{figure}

Fig.~\ref{fig:PSD} shows an informative illustration of the changes in the signal's time and frequency domain properties, and in its amplitude distribution, while it propagates through the signal processing chains. Specifically the properties with a linear chain (points (a), (b), and (c) in panel II of Fig.~\ref{fig:Linear}) and the ACDL (points I through V in Fig.~\ref{fig:Practical_Imp}) are highlighted. In Fig.~\ref{fig:PSD}, the black dashed lines correspond to the desired signal (without noise), and the colored solid lines correspond to the signal+noise mixtures based on the PRIME standard. The leftmost panels show the time domain traces, the rightmost panels show the PSDs, and the middle panels show the amplitude densities (PDFs). The value of parameter $\beta$ for Tukey's range is set to $\beta=3$. As it can be seen in the panels of row V, the difference signal largely reflects the temporal behavior and the amplitude of the noise. Thus, its output can be used to obtain the range for identifying the noise outliers (i.e., the clipping value $V_c/g$). From the panels of row II, it is clear that CMTF disproportionately affects signals with different temporal and/or amplitude structures and then reduces the spectral density of the impulsive noise in the signal passband without significantly affecting the signal of interest. The anti-aliasing (row III) and the baseband (row IV) filters further reduce the remaining noise to within the baseband, while the modified matched filter also compensates for the insertion of the CMTF in the signal chain. By comparing the panels of row (c) and row IV (specially PSDs panels), one can see the achieved improvement due to ACDL in the quality of the baseband signal is significant. In the following, we show the aforementioned improvement in terms of SNR and BER.

\begin{figure}[t]
\centering
\includegraphics[width=.5\textwidth,height=60mm]{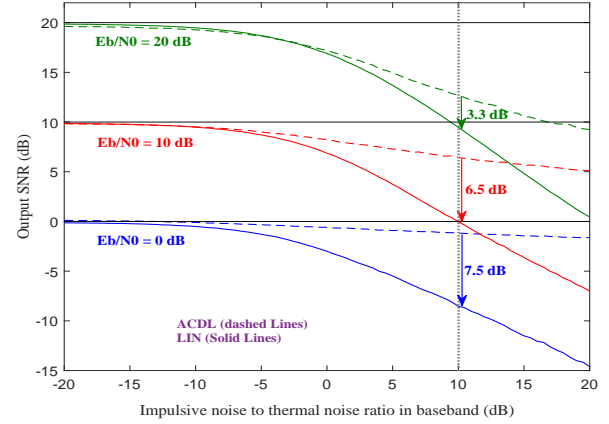}
\caption{Comparison of output SNR for the linear processing chain (solid lines) and ACDL (dashed lines). $1/\lambda=2e^{-5} s$.}
\label{fig:SNR}
\end{figure}

Fig.~\ref{fig:SNR} compares the output SNR performance for the linear processing chain and ACDL for various signal+noie compositions. As one can see in Fig.~\ref{fig:SNR}, for an effective value $\beta=3$, both linear and ACDL provide effectively equivalent performance when thermal noise dominates the impulsive noise. However, the ACDL shows its potency when the impulsive noise is dominant and in low SNR (SNR less than zero) its performance is insensitive to further increase in the impulsive noise. The robustness of the ACDL in different types of impulsive noise is demonstrated by considering the case when both asynchronous and cyclostationary impulsive noise impact the signal simultaneously. The BER performance of the ACDL for different values of SIR versus Eb/N0 is shown in Fig.~\ref{fig:BER vs SNR, fixed SIR}. The performance of the ACDL is compared with linear filter, blanking and clipping when the optimum thresholds for blanking and clipping are found based on an exhaustive numerical search. Fig.~\ref{fig:BER vs SNR, fixed SIR} shows that ACDL outperform other approaches, especially at high SNR.

\begin{figure}[t]
\centering
\includegraphics[width=.5\textwidth,height=60mm]{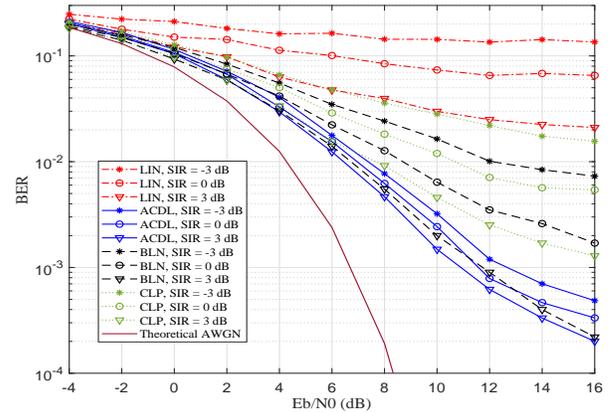}
\caption{BER versus Eb/N0 with fixed SIR. $1/\lambda=2e^{-5} s$.}
\label{fig:BER vs SNR, fixed SIR}
\end{figure}
\begin{figure}[t]
\centering
\includegraphics[width=.5\textwidth,height=60mm]{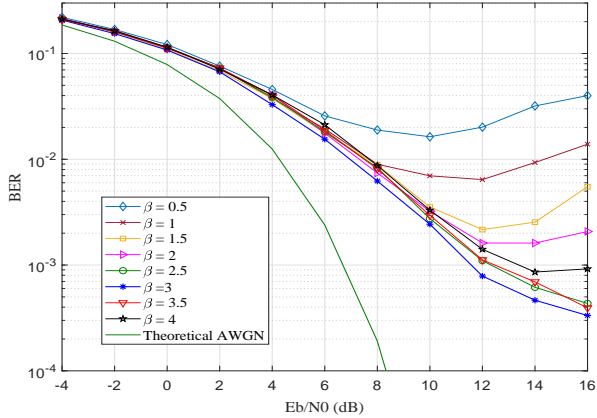}
\caption{Effect of $\beta$ on ACDL performance. SIR = 0 dB, $1/\lambda=2e^{-5} s$.}
\label{fig:Beta}
\end{figure}

It is important to mention that the range $[\alpha_-,\alpha_+]$ in Fig.~\ref{fig:BER vs SNR, fixed SIR} are determined by QTFs module and $\beta=3$ which is an effective value for range $\alpha$ but not the optimum one. It is clear that a fixed value of $\beta$ can not guarantee the optimum value of $\alpha$ for all kinds of noise, but an effective value of $\beta$ for a specific application can be easily found by training the ACDL in a short duration of time. The effect of $\beta$ on the performance of the ACDL is illustrated in Fig.~\ref{fig:Beta}. As it can be seen the value of $\beta$ is critical especially at high SNR but selecting a value near the optimum one (e.g., $\beta=2.5, 3.5$ in Fig.~\ref{fig:Beta}) can ensure a reasonable performance. Using inefficient $\beta$, i.e., with high deviation from the effective value, may cause considerable performance degradation at higher SNR. Such behavior is due to inappropriate elimination of the impulsive noise or cropping the desired signal in large or small $\beta$ values, respectively.

\section{Conclusion}

In this work, a practical implementation of adaptive analog nonlinear filter, referred to as Adaptive Canonical Differential Limiter (ACDL) is proposed to mitigate impulsive noise. The ACDL consists of two modules: Clipped Mean Tracking Filter (CMTF) and Quartile Tracking Filters (QTFs), which take care of outliers mitigation and finding a real-time range for parameter $\alpha$, respectively. In addition, a modified match filter is introduced to alleviate the effect of CMTF. We demonstrate the performance of the ACDL considering an OFDM-based PLC system with both asynchronous and cyclostationary impulsive noises. The results show that the ACDL can provide improvement in the overall signal quality ranging from distortionless behavior for low impulsive noise conditions to significant improvement in SNR or BER performance in the presence of a strong impulsive component. Moreover, the ACDL outperforms other approaches such as blanking and clipping in reducing the BER in impulsive noise environments. It is important to note that our filter can be deployed either as a stand-alone low-cost real-time solution for impulsive noise mitigation, or combined with other interference reduction techniques.

\bibliographystyle{IEEEtran}

\bibliography{IEEEabrv,Reference}

\end{document}